\documentclass[journal,twocolumn,10pt,final,letterpaper]{IEEEtran}

\usepackage{cite}
\usepackage{hyperref}
\ifCLASSINFOpdf
  \usepackage[pdftex]{graphicx}
    \graphicspath{{../figures/}}
\else
   \usepackage[dvips]{graphicx}
   \graphicspath{{../figures/}}
   \DeclareGraphicsExtensions{.eps}
\fi

\usepackage[cmex10]{amsmath}
\usepackage{amssymb,amsthm}
\usepackage{mathtools}

\newtheorem{remark}{Remark}
\newtheorem{lemma}{Lemma}
\newtheorem{theorem}{Theorem}
\newtheorem{corollary}{Corollary}
\newtheorem{definition}{Definition}

\newcommand{\hypergeom}[4]{{}_{2}\hspace{-.02cm}F_{\hspace{-.02cm}1}\hspace{-.1cm}\left(#1,#2,#3;#4\right)}

\newcommand{\cp}{\mathtt{P}_{\text{\textnormal{c}}}}

\newcommand{\sinr}{\mathtt{SINR}}
\newcommand{\rate}{\mathtt{R}}
\renewcommand{\th}{^\text{th}}

\newcommand{\mathdef}{~\raisebox{-0.03cm}{$\triangleq$}~}

\newcommand{\x}{\mathsf{x}}

\usepackage{array}
\usepackage{mdwmath}
\usepackage{mdwtab}
\usepackage{enumerate}
\usepackage{cases}

\usepackage[caption=false,font=footnotesize,labelfont=sf,textfont=sf]{subfig}

\usepackage{fixltx2e}
\usepackage{url}

\allowdisplaybreaks

\begin{document}

\title{Analysis of Heterogeneous Cellular Networks under Frequency Diversity and Interference Correlation}

\author{Ralph~Tanbourgi and Friedrich~K.~Jondral

\thanks{R.~Tanbourgi and F.~K.~Jondral are with the Communications Engineering Lab (CEL), Karlsruhe Institute of Technology (KIT), Germany. Email: \texttt{\{ralph.tanbourgi, friedrich.jondral\}@kit.edu}. This work was partially supported by the German Research Foundation (DFG) within the Priority Program 1397 "COIN" under grant No. JO258/21-2.}
}

\maketitle

\begin{abstract}
Analyzing heterogeneous cellular networks (HCNs) has become increasingly complex, particularly due to irregular base station locations, massive deployment of small cells, and flexible resource allocation. The latter is usually not captured by existing stochastic models for analytical tractability. In this work, we develop a more realistic stochastic model for a generic $K$-tier HCN, where users are served in the downlink under frequency diversity. We derive the rate coverage probability for this case, taking into account the interference correlation across different parts of the allocated spectrum. Our results indicate that ignoring this type of correlation may considerably overestimate the offered date rate. Furthermore, the gain of spectrum diversification is significant, particularly at low to moderate target data rates.
\end{abstract}


\IEEEpeerreviewmaketitle

\section{Introduction}\label{sec:introduction}
Flexible segmentation and resource allocation of potentially large transmission bandwidths is a prominent feature of 4G networks, and is considered no less important for 5G networks\cite{andrews_5G}. In Long Term Evolution (LTE), for instance, it is possible to assign non-contiguous spectrum to certain users using distributed virtual resource blocks (DVRBs)\cite{gosh10}. Carrier aggregation\cite{lteBook}, on the other side, allows users to operate on much larger bandwidths. Such techniques offer {\it frequency diversity} as the different parts of the allocated spectrum usually undergo independent fading. Besides, analyzing cellular networks has become increasingly complex, particularly due to their heterogeneity caused by the massive deployment of small cells within macro coverage areas. Existing works that analyze the performance of such heterogeneous cellular networks (HCNs), e.g.,\cite{singh13}, mostly assume that users are served on {\it purely coherent} resources. Consequently, these works do not capture non-contiguous spectrum allocation nor allocations larger than the coherence bandwidth, and hence they cannot capture the effect of frequency diversity. But analyzing this effect is challenging since the experienced interference is usually correlated across different parts of the spectrum due to the common locations of interfering base stations (BSs). This type of correlation was considered in\cite{lin13}, where the {\it ergodic} rate with carrier aggregation was studied. To obtain a better understanding of frequency diversity in HCNs, one has to look at the rate {\it distribution}, which is the main focus of this paper.

{\bf Contributions and Outcomes:}
In this work, we develop a tractable while realistic model for a generic $K$-tier HCN and study the rate distribution for a typical user in the downlink under frequency diversity and spatial interference correlation. We model the BS locations by a Poisson point process (PPP)\cite{stoyan95,HaenggiBook} and assume a two-block independent-fading model for capturing the basic effect of frequency diversity, cf. Section~\ref{sec:model}. Our results indicate that, depending on the degree of spectrum diversification, rate gains of 40--90\% are obtained in a typical three-tier HCN scenario. When spatial interference correlation is ignored, the offered rate is overestimated by 3--6\%.

{\bf Notation:} We use sans-serif-style letters ($\mathsf{z}$) and serif-style letters ($z$) for denoting random variables and their realizations or variables, respectively. We define $(z)^{+}\mathdef\max\{0,z\}$.

\begin{figure}[t]
	\centering
	\includegraphics[width=.48\textwidth]{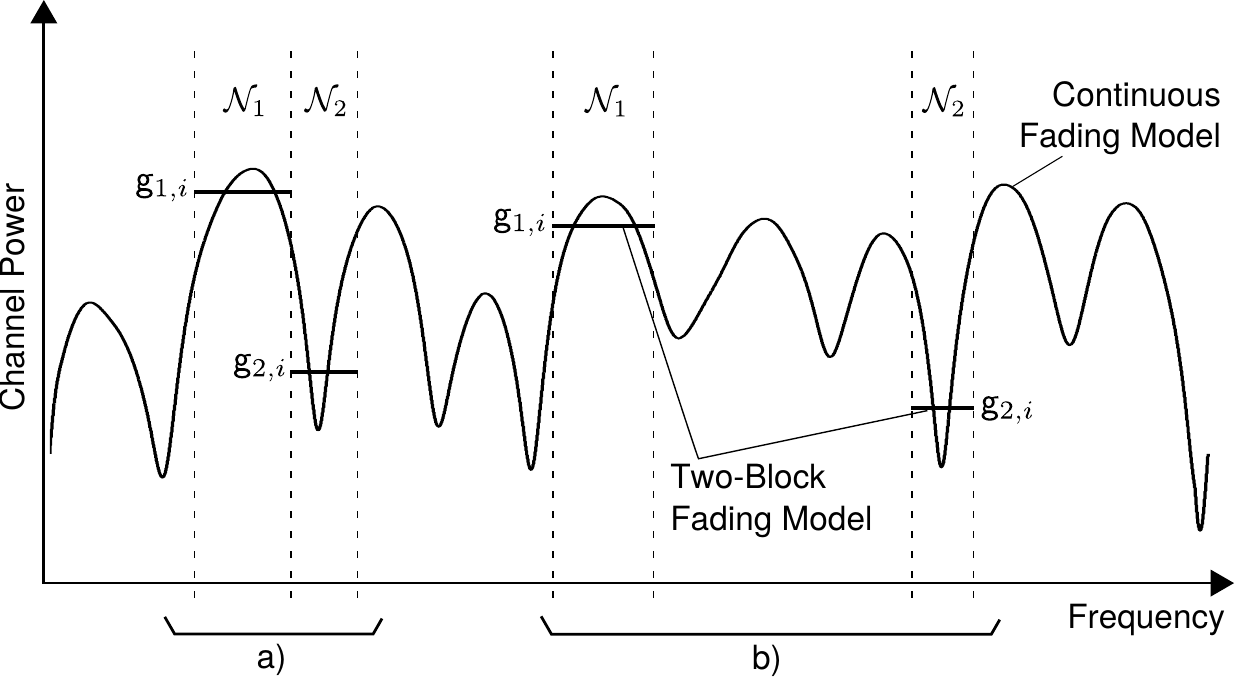}
	\caption{Two-block independent-fading model: a) allocated contiguous spectrum exceeds coherence bandwidth; b) non-contiguous spectrum allocation.}
	\label{fig:fading_model}
\end{figure}

\section{System Model}\label{sec:model}
We consider a generic $K$-tier HCN with single-antenna BSs in the downlink. The BS locations of each tier are modeled by an independent stationary planar PPP $\Phi_k$ with average density $\lambda_k$. We assume that single-antenna users are scattered in the plane according to a stationary planar PPP. By Slivnyak's Theorem\cite{stoyan95,HaenggiBook}, we can focus our analysis on a typical user located in the origin $o$. At the typical user, the long-term signal power received from a $k\th$ tier BS located at $\x_i\in\Phi_k$ is $P_k\|\x_i\|^{-\alpha_k}$, where $P_k$ is the tier-specific BS transmit power and $\|\cdot\|^{-\alpha_k}$ is the distance-dependent path loss with exponent $\alpha_k>2$. Users associate with the BS providing the highest long-term received power. The location of the BS serving the typical user is hence given by the point from $\Phi\mathdef\cup_{k=1}^{K}\Phi_k$ associated with the Voronoi cell covering the origin $o$, cf.\cite{singh13}. Assuming that the serving BS is from tier $\ell$, we label its location by $\x_o\in\Phi_{\ell}$ without loss of generality. For convenience, we define $\Phi^{o}\mathdef\Phi\setminus\{\x_o\}$ (similarly, $\Phi_k^{o}\mathdef\Phi_k\setminus\{\x_o\}$), i.e., the set of interfering BSs.

\newcounter{mycounter}
\begin{figure*}[!t]
\normalsize
\setcounter{mycounter}{\value{equation}}
\setcounter{equation}{5}
\begin{IEEEeqnarray}{rCl}	\cp=-\sum\limits_{\ell=1}^{K}2\pi\lambda_\ell\int_{0}^{\infty}\hspace{-.1cm}\int_{0}^{\infty}\hspace{-.1cm}y\, e^{-\frac{\beta(z)\sigma^2}{P_{\ell}y^{-\alpha_{\ell}}}}\frac{\mathrm d}{\mathrm dw}\hspace{-.1cm}\left[\exp\left(-\frac{\gamma(w)\sigma^2}{P_{\ell}y^{-\alpha_{\ell}}}-\pi\sum\limits_{k=1}^{K}\lambda_k\hat P_k^{2/\alpha_k}y^{2/\hat\alpha_k}\Big(1+\mathcal{F}\left[\beta(z),\gamma(w),\alpha_k\right]\Big)\right)\right]_{w=z}\hspace{-.5cm}\mathrm dy\,\mathrm dz\IEEEeqnarraynumspace\label{eq:main}
\end{IEEEeqnarray}
\setcounter{equation}{0}
\hrulefill
\end{figure*}

BS $\x_o$ uses $N$ orthogonal (frequency) resources for serving the typical user.\footnote{The analysis does not change when considering time-frequency resources.} It is reasonable to assume that the channel fading on each resource is frequency-flat. Yet some resources may experience a different fading realization, e.g., if the total allocated spectrum exceeds the coherence bandwidth or if the allocation is non-contiguous, cf. Fig.~\ref{fig:fading_model}. To capture this effect while maintaining analytical tractability, we assume the following two-block fading model. A possible extension toward a general $M$-block model is deferred to future work.

{\bf Two-Block Independent-Fading Model:} We partition the set of resources allocated to the typical user into two subsets $\mathcal{N}_1$ and $\mathcal{N}_2$ such that (i) all resources falling into one subset experience the {\it same} fading realization, (ii) while the fading is considered {\it independent} across the two subsets. For a given BS located at $x_i$ operating on any resource from $\mathcal{N}_1$ ($\mathcal{N}_2$), the corresponding signal is thus received by the typical user under a common fading gain $\mathsf{g}_{1,i}$ ($\mathsf{g}_{2,i}$). By assumption (ii) $\mathsf{g}_{1,i}$ and $\mathsf{g}_{2,i}$ are statistically independent. 
See Fig.~\ref{fig:fading_model} for an illustration.

We further assume $\mathsf{g}_{m,i}\sim\text{Exp}(1)$ for $m=1,2$ and all $i\in\mathbb{N}_{0}$ (Rayleigh fading), and that the $\mathsf{g}_{m,i}$ are statistically independent across BSs.  We denote by $N_1$ and $N_2$ the cardinality of $\mathcal{N}_1$ and $\mathcal{N}_2$, respectively, and hence $N_1+N_2=N$. Note that the frequently used case of fully coherent resources can be recovered by setting either $N_1$ or $N_2$ equal to zero.

The next lemma gives the probability that a user associates with the $\ell\th$ tier as well as the conditional probability density function (PDF) of the distance to the serving BS.
\begin{lemma}[Association Probability and Distance PDF\cite{singh13}]\label{lem:asso}
A user associates with the $\ell\th$ tier with probability
\begin{IEEEeqnarray}{rCl}
   A_{\ell}=2\pi\lambda_\ell\int_{0}^{\infty}y\,\exp\left(-\pi\sum\limits_{k=1}^{K}\lambda_k \hat{P}_k^{2/\alpha_k}y^{2/\hat{\alpha}_k}\right)\mathrm dy,
\end{IEEEeqnarray}
where $\hat{P}_{k}\triangleq P_k/P_\ell$ and $\hat{\alpha}_k\mathdef\alpha_k/\alpha_\ell$. The PDF of the distance $\mathsf{y}\mathdef\|\x_o\|$ to the serving BS, given that it belongs to tier $\ell$, is
\begin{IEEEeqnarray}{rCl}
   f_{\mathsf{y},\ell}(y)=\frac{2\pi\lambda_\ell y}{A_{\ell}}\exp\left(-\pi\sum\limits_{k=1}^{K}\lambda_k \hat{P}_k^{2/\alpha_k}y^{2/\hat{\alpha}_k}\right),\quad y\geq0.\IEEEeqnarraynumspace
\end{IEEEeqnarray}
\end{lemma}


\section{Rate Coverage Probability Analysis}
The rate offered to the typical user is determined by $N_1$, $N_2$ and by the signal-to-interference-plus-noise ratio ($\sinr$) experienced on the corresponding resources. Given that the typical user associates with an $\ell\th$-tier BS at distance $y$, the resulting $\sinr$ on any resource from $\mathcal{N}_m$ is
\begin{IEEEeqnarray}{rCl}
   \sinr_{m,\ell}(y)\mathdef\frac{\mathsf{g}_{m,o}\,P_{\ell}\,y^{-\alpha_\ell}}{\sum_{k=1}^{K}\mathsf{I}_{m,k}+\sigma^2},\label{eq:sinr}
\end{IEEEeqnarray}
where $\mathsf{I}_{m,k}\mathdef\sum_{\x_i\in\Phi_k^{o}}\mathsf{g}_{m,i}P_k\|\x_i\|^{-\alpha_k}$ is the other-cell interference and $\sigma^{2}$ is the AWGN power. The total rate conveyed to the typical user during a single downlink phase then becomes
\begin{IEEEeqnarray}{rCl}
   \rate_{\ell}\mathdef\rate_{1,\ell}+\rate_{2,\ell}&=&\sum\limits_{m=1}^{2}N_m\log_2(1+\sinr_{m,\ell}),\label{eq:rate_def}
\end{IEEEeqnarray}
where $\rate_{m,\ell}$ denotes the product of link spectral efficiency corresponding to a channel realization from the subset $\mathcal{N}_m$ and the number of resources corresponding to the subset $\mathcal{N}_m$, given that the serving BS is from tier $\ell$. Although the individual $\rate_{m,\ell}$ in \eqref{eq:rate_def} comprise only statistically independent quantities, they are correlated across $m$ due to common locations of interfering BSs. Hence, $\rate_{\ell}$ is a sum of correlated random variables whose correlation structure is difficult to characterize.

\begin{definition}[Rate Coverage Probability]
The rate coverage probability for the typical user is defined as
\begin{IEEEeqnarray}{rCl}
   \cp\mathdef\mathbb{P}\left(\rate\geq \tau\right), \quad \tau\geq0.\label{eq:cp_def}
\end{IEEEeqnarray}
\end{definition}
Besides its definition as a statistical quantity, $\cp$ can be also interpreted as the average fraction of users in the network covered by a rate $\rate$ no less than a given threshold $\tau$\cite{singh13}. Given the aforementioned opaque correlation structure inherent to $\rate_{\ell}$, the challenging nature of deriving the $\cp$ may now become apparent. This task is addressed in the following Theorem.

\begin{theorem}[Main Result]\label{thm:main}
	The coverage probability for the typical user in the described setting is given in \eqref{eq:main} at the top of the page, where $\mathcal{F}[a,b,c]$ is given in \eqref{eq:proof1functional}, $\beta(z)\mathdef2^{(\tau-z)^{+}/N_{1}}-1$, and $\gamma(w)\mathdef2^{w/N_2}-1$.
\end{theorem}
\setcounter{equation}{6}
\begin{IEEEproof}
   See Appendix.
\end{IEEEproof}
As can be seen in \eqref{eq:main}, Theorem~\ref{thm:main} is given as a double integral, which can be easily solved using standard numerical software. We recover the rate coverage probability result from\cite{singh13} by setting $N_{1}$ or $N_{2}$ equal to zero. The computation complexity can be further reduced as discussed next. 

\begin{figure*}[!t]
\centerline{\subfloat[]{
	\includegraphics[width=0.48\textwidth]{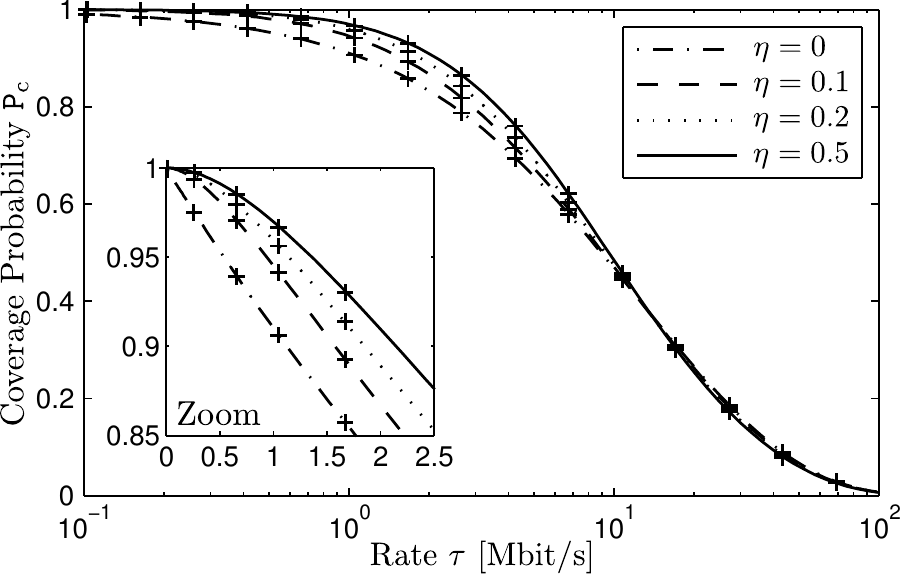}
	\label{fig:cp}}
	\hfil
	\subfloat[]{
	\includegraphics[width=0.508\textwidth]{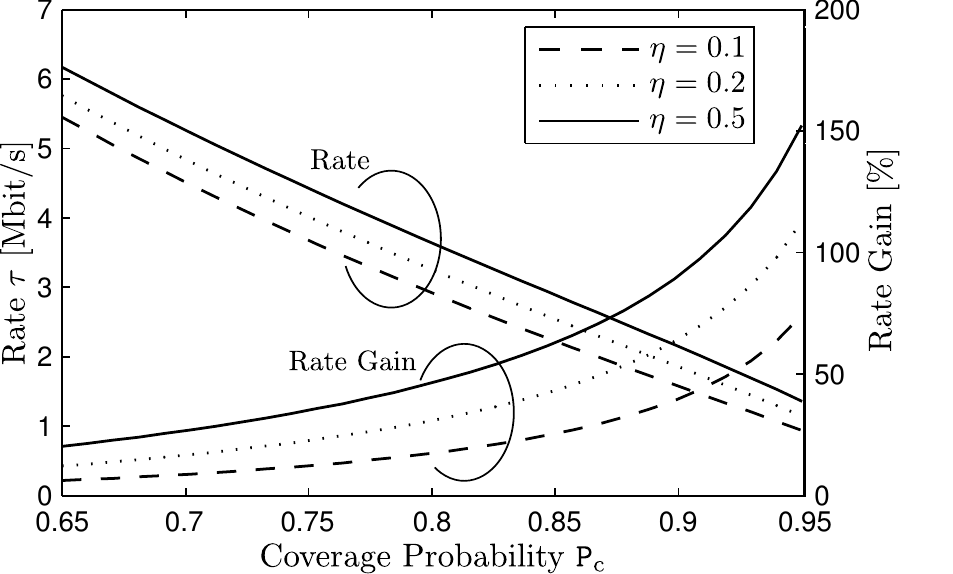}
	\label{fig:gain}}}
	\caption{a) Coverage probability for different $\eta$. Marks represent simulation results. b) Rate and rate gain relative to the case $\eta=0$ for different $\cp$ and $\eta>0$.}\vspace{-.2cm}
\end{figure*}

Due to the exponential form of the integrand, the expression inside the exponent in \eqref{eq:main} is the relevant part regarding the $\mathrm d/\mathrm dw$ operation. 
\begin{remark}
   The differentiation $\mathrm d/\mathrm dw\mathcal{F}[\beta(z),\gamma(w),\alpha_k]$ at $w=z$ can be calculated as
   \begin{IEEEeqnarray}{rCl}
		&&\frac{\mathrm d}{\mathrm dw}\mathcal{F}[\beta(z),\gamma(w),\alpha_k]\Big|_{w=z}=\frac{\beta(z)\,2^{z/N_{2}}\log2}{N_{2}(\beta(z)-\gamma(z))^2}\IEEEnonumber\\
		&&\qquad\times\left[\Psi(\beta(z),\alpha_k)-\left(1-\frac{\gamma(z)}{\beta(z)}\right)\frac{2/\alpha_k}{1+\gamma(z)^{-1}}\right.\IEEEnonumber\\
		&&\quad\qquad\quad\left.-\left(1+\frac{2}{\alpha_k}\left(1-\frac{\gamma(z)}{\beta(z)}\right)\right)\Psi(\gamma(z),\alpha_k)\right],
   \end{IEEEeqnarray}
   where $\Psi(p,q)$ is given by \eqref{eq:proof1step7}.
\end{remark}
It is possible to express $\Psi(p,q)$ in terms of the Gaussian hypergeometric function, i.e., $\Psi(p,q)=2\pi p^{2/q}\csc(2\pi/q)/q-\hypergeom{1}{2/q}{1+2/q}{-1/p}$. With \cite{olver10,wolfram_hyper} $\hypergeom{\cdot}{\cdot}{\cdot}{\cdot}$ can in some cases be expressed through elementary functions, for instance when $\alpha_k=4$,
\begin{IEEEeqnarray}{rCl}
   \mathcal{F}[\beta(z),\gamma(w),4]&=&\frac{\beta(z)^{3/2}\arctan\sqrt{\beta(z)}}{\beta(z)-\gamma(w)}\IEEEnonumber\\
   &&-\frac{\gamma(w)^{3/2}\arctan\sqrt{\gamma(w)}}{\beta(z)-\gamma(w)}.\IEEEeqnarraynumspace
\end{IEEEeqnarray}

Densely deployed cellular networks typically operate in the interference-limited regime\cite{goldsmith05}, where interference dominates noise. If, in addition, the path loss law does not differ significantly across tiers, then the coverage probability result from Theorem~\ref{thm:main} reduces to the following single-integral form.
\begin{corollary}[No Noise, Equal Path Loss]
   With no noise ($\sigma^2=0$) and equal path loss ($\alpha_k=\alpha$), \eqref{eq:main} becomes
   \begin{IEEEeqnarray}{rCl}
      \cp=-\int_{0}^{\infty}\frac{\mathrm d}{\mathrm dw}\left[\frac{1}{\mathcal{F}\left[\beta(z),\gamma(w),\alpha\right]}\right]_{w=z}\mathrm dz.\IEEEeqnarraynumspace
   \end{IEEEeqnarray}
\end{corollary}

\section{Numerical Results and Discussion}\label{sec:numerical}

In this section, we discuss through numerical examples how the developed model can be used for network analysis. In particular, we study the effect of spatial correlation due to common BS locations on the rate coverage probability. We assume a three-tier scenario ($K=3$) with macro, pico, and femto BSs, and with typical tier-specific parameters similar to\cite{3gpp_tr_36814,gosh12}, cf. Table~\ref{tab:param}. We further set $N=49\times180$ kHz, which is comparable to 7 resource blocks assuming the LTE standard\cite{gosh10}. Receiver noise is set to $\sigma^2=-104$ dBm.

{\bf Effect of Spectrum Diversification:} Fig.~\ref{fig:cp} shows the rate coverage probability $\cp$ for $\eta=[0;0.1;0.2;0.5]$, where $\eta\mathdef N_{1}/N$ is the fraction of total resources spanned by $\mathcal{N}_{1}$. First, it can be seen that the theoretic results perfectly match the simulation results, which were obtained by averaging over $10^{5}$ network realizations. Furthermore, the coverage probability increases as more diverse spectrum is occupied, with the maximum increase attained at $\eta=0.5$ (equal split between $\mathcal{N}_1$ and $\mathcal{N}_2$). This so-called frequency-diversity gain is exploited in non-contiguous resource allocation, e.g., DVRB allocation in LTE\cite{gosh10}. Fig.~\ref{fig:gain} shows the gain in terms of offered rate $\tau$ relative to $\eta=0$ for different $\cp$ and $\eta>0$. For typical values of $\cp$, e.g., around 90\% of covered users, the rate gain ranges from roughly 40\% for $\eta=0.1$ to 90\% for $\eta=0.5$. For very low $\cp$, frequency diversity becomes slightly unfavorable.

{\bf Effect of Spatial Interference Correlation:} Recall that assuming purely coherent resources yields a pessimistic performance prediction for systems with non-contiguous resource allocation or large transmission bandwidths, see Fig.~\ref{fig:gain}. In contrast, assuming that the $\sinr_{m,\ell}$ in \eqref{eq:rate_def} are independent across $m$, e.g., for analytical tractability, yields an optimistic prediction as the correlation between the $\mathsf{I}_{m,k}$ due to common locations of interfering BSs is ignored. A corresponding $\cp$ expression for this case can be obtained by decomposing the expectation in \eqref{eq:proof1step5} into the product of two expectations over independent point sets for every tier. After invoking the moment generating function for PPPs, see Appendix, we obtain a new $\mathcal{F}[a,b,c]=\Psi(a,c)+\Psi(b,c)$ for this case.

\begin{table}[t]
\small
\renewcommand{\arraystretch}{1.3}
\caption{Tier-Specific Parameters used for Numerical Examples}
\label{tab:param}
\centering
\begin{tabular}{c||c|c|c}
\hline
Parameter	&	Tier 1 & Tier 2	&	Tier 3\\
\hline
BS density $\lambda_k$	& $4\,\text{BS/km}^2$ &$16\,\text{BS/km}^2$ &$40\,\text{BS/km}^2$\\
BS power $P_k$ & $46\,\text{dBm}$	& $30\,\text{dBm}$ &$24\,\text{dBm}$\\
Path loss $\alpha_k$	&	3.76 & 3.67 & 3.5\\
\hline 
\end{tabular}
\end{table}

\begin{figure}[t!]
   \centering
	\includegraphics[width=.48\textwidth]{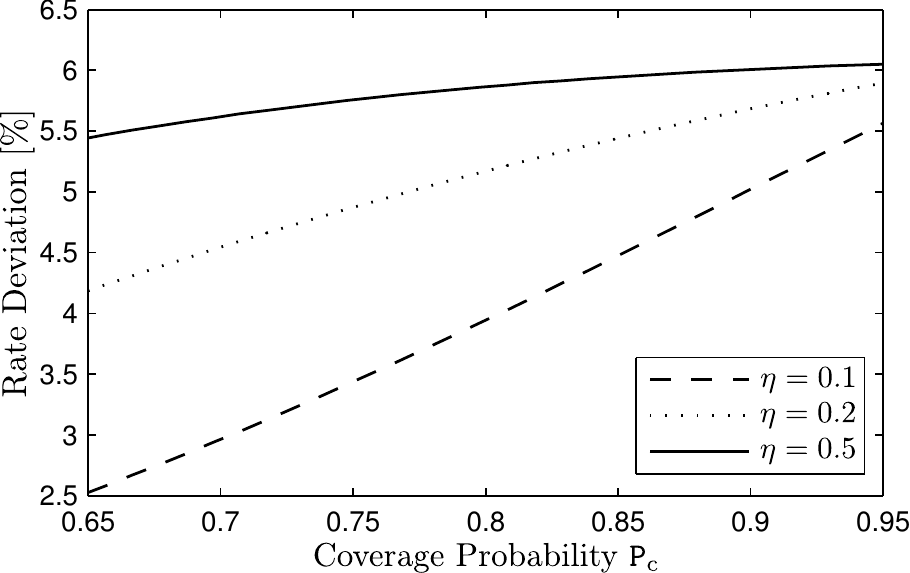}
	\caption{Rate deviation when spatial interference correlation is ignored for different $\cp$ and $\eta>0$.}
	\label{fig:dev}
\end{figure}

Equipped with Theorem~\ref{thm:main}, we next analyze the impact of spatial interference correlation on the system performance by comparing the exact $\cp$ to the $\cp$ when correlation is ignored. Fig.~\ref{fig:dev} shows the rate deviation caused by ignoring correlation. It can be seen that, depending on the degree of diversification, the rate is overestimated by 3--6\%. For $\eta=0.5$ and $\cp=0.9$ for instance, the absolute rate difference is $129$ kbit/s.

\section{Conclusion}
We developed a tractable while realistic stochastic model for a generic $K$-tier HCN, where users are served in the downlink under frequency diversity. This type of diversity could be due to a transmission bandwidth larger than the coherence bandwidth or due to non-contiguous spectrum allocation. Assuming a two-block independent-fading model, we derived the rate coverage probability for a typical user, taking into account the interference correlation across the different parts of the allocated spectrum due to common locations of interfering BSs. Several practical insights followed from our theoretic results. Possible future work includes an extension of the two-block fading model to a more general $M$-block fading model.

\appendix

\section{Proof of Theorem~\ref{thm:main}}
Using Lemma~\ref{lem:asso} and the law of total probability in \eqref{eq:cp_def} yields
\begin{IEEEeqnarray}{rCl}
	\cp&=&\sum\limits_{\ell=1}^{K}A_{\ell}\int_{0}^{\infty}f_{\mathsf{y},\ell}(y)\,\mathbb{P}(\rate_{\ell}(y)\geq\tau)\,\mathrm dy.\label{eq:proof1step1}
\end{IEEEeqnarray}
Next, we treat the conditional coverage probability $\mathbb{P}(\rate_{\ell}(y)\geq\tau)$. After conditioning this term on $\Phi^{o}$ to remove the dependency between $\rate_{1,\ell}$ and $\rate_{2,\ell}$, this term can be written as
\begin{IEEEeqnarray}{rCl}
	&&\mathbb{P}(\rate_{\ell}(y)\geq\tau)
	=\mathbb{E}_{\Phi^{o}}\left[\mathbb{P}(\rate_{1,\ell}(y)\geq\tau-\rate_{2,\ell}(y)|\Phi^{o})\right]\IEEEnonumber\\
	&&\quad=\mathbb{E}_{\Phi^{o}}\hspace{-.1cm}\left[\int_{0}^{\infty}\mathbb{P}(\rate_{1,\ell}(y)\geq\tau-z|\Phi^{o})\,f_{\rate_{2,\ell}(y)|\Phi^{o}}(z)\,\mathrm dz\right].\IEEEeqnarraynumspace\label{eq:proof1step2}
\end{IEEEeqnarray}
The first term inside the integral in \eqref{eq:proof1step2} can be written as
\begin{IEEEeqnarray}{rCl}
	&&\mathbb{P}(\rate_{1,\ell}(y)\geq\tau-z|\Phi^{o})\overset{\text{(a)}}{=}\mathbb{P}\left(\sinr_{1,\ell}\geq\beta(z)|\Phi^{o}\right)\IEEEnonumber\\
	&&\quad=\mathbb{P}\left(\mathsf{g}_{1,o}\geq\frac{\beta(z)}{P_{\ell}y^{-\alpha_{\ell}}}\left\{\sum_{k=1}^{K}\mathsf{I}_{1,k}+\sigma^2\right\}\middle|\Phi^{o}\right)\IEEEeqnarraynumspace\IEEEnonumber\\
	&&\quad\overset{\text{(b)}}{=}e^{-\frac{\beta(z)\sigma^2}{P_{\ell}y^{-\alpha_{\ell}}}}\prod_{k=1}^{K}\prod_{x_i\in\Phi_{k}^{o}}\mathbb{E}\left[e^{-\beta(z)\hat P_{\ell}y^{\alpha_{\ell}}\|x_i\|^{-\alpha_{k}}\mathsf{g}_{1,i}}\middle|\Phi_{k}^{o}\right]\IEEEnonumber\\
	&&\quad\overset{\text{(c)}}{=}e^{-\frac{\beta(z)\sigma^2}{P_{\ell}y^{-\alpha_{\ell}}}}\prod_{k=1}^{K}\prod_{x_i\in\Phi_{k}^{o}}\frac{1}{1+\beta(z)\hat P_{k}y^{\alpha_{\ell}}\|x_i\|^{-\alpha_{k}}},\label{eq:proof1step3}
\end{IEEEeqnarray}
where we have substituted $\beta(z)\mathdef2^{(\tau-z)^{+}/N_{1}}-1$ in (a), (b) follows from evaluating the probability with respect to $\mathsf{g}_{1,o}$ and noting that all remaining gains are i.i.d. across all BSs, and (c) is obtained by using the fact that $\mathbb{E}[e^{-s\mathsf{g}}]=1/(1+s)$ for $\mathsf{g}\sim\text{Exp}(1)$. Using the same approach, the second term inside the integral in \eqref{eq:proof1step2} can be written as
\begin{IEEEeqnarray}{rCl}
	&&f_{\rate_{2,\ell}(y)|\Phi^{o}}(z)=\frac{\mathrm d}{\mathrm dw}\Big[\mathbb{P}\left(\rate_{2,\ell}(y)\leq w\right)\Big]_{w=z}\IEEEnonumber\\
	&&=-\frac{\mathrm d}{\mathrm dw}\hspace{-.1cm}\left[e^{-\frac{\gamma(w)\sigma^2}{P_{\ell}y^{-\alpha_{\ell}}}}\prod_{k=1}^{K}\prod_{x_i\in\Phi_{k}^{o}}\hspace{-.1cm}\frac{1}{1+\gamma(w)\hat P_{k}y^{\alpha_{\ell}}\|x_i\|^{-\alpha_{k}}}\right]_{w=z}\hspace{-.55cm},\label{eq:proof1step4}\IEEEeqnarraynumspace
\end{IEEEeqnarray}
where $\gamma(w)\mathdef2^{w/N_2}-1$.	Now observe that the expectation $\mathbb{E}_{\Phi^{o}}$ can be moved inside the integral over $z$ in \eqref{eq:proof1step2} by Fubini's Theorem\cite{gut05}. By Leibniz integration rule for infinite integrals\cite{olver10}, the differentiation $\mathrm d/\mathrm dw$ in \eqref{eq:proof1step4} can be moved outside the expectation $\mathbb{E}_{\Phi^{o}}$. Exploiting the independence property of the $\Phi_k$, \eqref{eq:proof1step2} can hence be rewritten as
\begin{IEEEeqnarray}{rCl}
	&&\mathbb{P}(\rate_{\ell}(y)\geq\tau)=-\int_{0}^{\infty}e^{-\frac{\beta(z)\sigma^2}{P_{\ell}y^{-\alpha_{\ell}}}}\frac{\mathrm d}{\mathrm dw}\left[e^{-\frac{\gamma(w)\sigma^2}{P_{\ell}y^{-\alpha_{\ell}}}}\right.\IEEEnonumber\\
	&&\qquad\times\prod_{k=1}^{K}\mathbb{E}_{\Phi_k^{o}}\left[\prod_{x_i\in\Phi_{k}^{o}}\frac{1}{1+\beta(z)\hat P_{k}y^{\alpha_{\ell}}\|\x_i\|^{-\alpha_{k}}}\right.\IEEEnonumber\\
	&&\left.\left.\qquad\qquad\qquad\qquad\times\frac{1}{1+\gamma(w)\hat P_{k}y^{\alpha_{\ell}}\|\x_i\|^{-\alpha_{k}}}\right]\right]_{w=z}\hspace{-.2cm}\mathrm dz.\IEEEeqnarraynumspace\label{eq:proof1step5}
\end{IEEEeqnarray}
After invoking the moment generating function for PPPs\cite{stoyan95}, expanding $\tfrac{1}{1+\beta(z)\ldots}\tfrac{1}{1+\gamma(w)\ldots}$ using partial-fraction decomposition as done in\cite{tanbourgi14_1}, and further algebraic manipulations, the second line in \eqref{eq:proof1step5} becomes
\begin{IEEEeqnarray}{rCl}
	&&\exp\left(-\pi\sum\limits_{k=1}^{K}\lambda_k\hat P_k^{2/\alpha_k}y^{2/\hat\alpha_k}\mathcal{F}[\beta(z),\gamma(w),\alpha_k]\right),\IEEEeqnarraynumspace\label{eq:proof1step6}
\end{IEEEeqnarray}
where the functional $\mathcal{F}[a,b,c]$ is defined as
\begin{IEEEeqnarray}{rCl}
	\mathcal{F}[a,b,c]\mathdef \frac{a\,\Psi(a,c)}{a-b}-\frac{b\,\Psi(b,c)}{a-b},\IEEEeqnarraynumspace\label{eq:proof1functional}
\end{IEEEeqnarray}
and
\begin{IEEEeqnarray}{rCl}
	\Psi(p,q)\mathdef \int_{1}^{\infty}\frac{p}{p+t^{q/2}}\mathrm dt.\label{eq:proof1step7}
\end{IEEEeqnarray}
Combining \eqref{eq:proof1step1}, \eqref{eq:proof1step2} and \eqref{eq:proof1step5}--\eqref{eq:proof1step7} yields the result.



\begin{thebibliography}{10}
\providecommand{\url}[1]{#1}
\csname url@samestyle\endcsname
\providecommand{\newblock}{\relax}
\providecommand{\bibinfo}[2]{#2}
\providecommand{\BIBentrySTDinterwordspacing}{\spaceskip=0pt\relax}
\providecommand{\BIBentryALTinterwordstretchfactor}{4}
\providecommand{\BIBentryALTinterwordspacing}{\spaceskip=\fontdimen2\font plus
\BIBentryALTinterwordstretchfactor\fontdimen3\font minus
  \fontdimen4\font\relax}
\providecommand{\BIBforeignlanguage}[2]{{%
\expandafter\ifx\csname l@#1\endcsname\relax
\typeout{** WARNING: IEEEtran.bst: No hyphenation pattern has been}%
\typeout{** loaded for the language `#1'. Using the pattern for}%
\typeout{** the default language instead.}%
\else
\language=\csname l@#1\endcsname
\fi
#2}}
\providecommand{\BIBdecl}{\relax}
\BIBdecl

\bibitem{andrews_5G}
J.~G. Andrews \emph{et~al.}, ``What will {5G} be?'' vol.~PP, no.~99, pp. 1--1,
  2014.

\bibitem{gosh10}
A.~Ghosh, J.~Zhang, J.~G. Andrews, and R.~Muhamed, \emph{Fundamentals of LTE},
  1st~ed.\hskip 1em plus 0.5em minus 0.4em\relax Upper Saddle River, NJ, USA:
  Prentice Hall Press, 2010.

\bibitem{lteBook}
S.~Sesia, I.~Toufik, and M.~Baker, \emph{{LTE - The UMTS Long Term Evolution:
  From Theory to Practice}}, 2nd~ed.\hskip 1em plus 0.5em minus 0.4em\relax
  Wiley, Sep. 2011.

\bibitem{singh13}
S.~Singh, H.~S. Dhillon, and J.~G. Andrews, ``Offloading in heterogeneous
  networks: Modeling, analysis, and design insights,'' \emph{{IEEE} Trans.
  Wireless Commun.}, vol.~12, no.~5, pp. 2484--2497, Dec. 2013.

\bibitem{lin13}
X.~Lin, J.~G. Andrews, and A.~Ghosh, ``Modeling, analysis and design for
  carrier aggregation in heterogeneous cellular networks,'' \emph{{IEEE} Trans.
  Commun.}, vol.~61, no.~9, pp. 4002--4015, Sep. 2013.

\bibitem{stoyan95}
D.~Stoyan, W.~Kendall, and J.~Mecke, \emph{Stochastic Geometry and its
  Applications}, 2nd~ed.\hskip 1em plus 0.5em minus 0.4em\relax Wiley, 1995.

\bibitem{HaenggiBook}
M.~Haenggi, \emph{Stochastic Geometry for Wireless Networks}.\hskip 1em plus
  0.5em minus 0.4em\relax Cambridge University Press, 2012.

\bibitem{olver10}
F.~W. Olver \emph{et~al.}, \emph{NIST Handbook of Mathematical Functions},
  1st~ed.\hskip 1em plus 0.5em minus 0.4em\relax New York, NY, USA: Cambridge
  University Press, 2010.

\bibitem{wolfram_hyper}
\BIBentryALTinterwordspacing
{Wolfram}. {Gauss Hypergeometric Function}. [Online]. Available:
  \url{http://functions.wolfram.com/HypergeometricFunctions/Hypergeometric2F1/}
\BIBentrySTDinterwordspacing

\bibitem{goldsmith05}
A.~Goldsmith, \emph{Wireless Communications}.\hskip 1em plus 0.5em minus
  0.4em\relax Cambridge, New York: Cambridge University Press, 2005.

\bibitem{3gpp_tr_36814}
3GPP, ``Further advancements for {E-UTRA},'' TR 36.814, Tech. Rep., Mar. 2009.

\bibitem{gosh12}
A.~Ghosh \emph{et~al.}, ``Heterogeneous cellular networks: From theory to
  practice,'' \emph{{IEEE} Commun. Mag.}, vol.~50, no.~6, pp. 54--64, Jun.
  2012.

\bibitem{gut05}
A.~Gut, \emph{Probability: A Graduate Course}, ser. Springer Texts in
  Statistics.\hskip 1em plus 0.5em minus 0.4em\relax New York, NY: Springer,
  2005.

\bibitem{tanbourgi14_1}
R.~Tanbourgi, H.~S. Dhillon, J.~G. Andrews, and F.~K. Jondral, ``Dual-branch
  {MRC} receivers in the cellular downlink under spatial interference
  correlation,'' in \emph{20th European Wireless Conference}, May 2014, pp.
  1--6.

\end{thebibliography}
\end{document}